\documentclass[pra,aps]{revtex4}
\newcommand{\be}{\begin{equation}}
\newcommand{\ee}{\end{equation}}
\newcommand{\br}{\begin{eqnarray}}
\newcommand{\er}{\end{eqnarray}}
\newcommand{\nn}{\nonumber}
\newcommand{\bd}{\begin{displaymath}}
\newcommand{\ed}{\end{displaymath}}

\newcommand{\bib}{\bibitem}
\newcommand{\bfig}{\begin{figure}}
\newcommand{\efig}{\end{figure}}
\usepackage{ifvtex}
\usepackage[dvips]{graphicx}
\usepackage{amsmath}
\usepackage{amssymb}
\usepackage{eucal}

\def\eps{\epsilon}
\def\rpar{\right)}
\def\lpar{\left(}
\def\rbk{\right]}
\def\lbk{\left[}
\def\rbr{\right\}}
\def\lbr{\left\{}
\def\lb{\label}

\def\tr{\mbox{${\rm Tr}$}}
\def\ro{\mbox{\boldmath $\rho$}}

\def\re{\mbox{${\rm Re}$}}
\def\rg{\rangle}
\def\lg{\langle}

\begin{document}
\title{Extended Cahill-Glauber formalism for finite-dimensional spaces: I. Fundamentals}
\author{M. Ruzzi and D. Galetti}
\affiliation{Instituto de F\'{\i}sica Te\'{o}rica, Universidade Estadual Paulista, \\
             Rua Pamplona 145, 01405-900, S\~{a}o Paulo, SP, Brazil \\
             E-mail address: mruzzi@ift.unesp.br and galetti@ift.unesp.br}
\author{M. A. Marchiolli}
\affiliation{Instituto de F\'{\i}sica de S\~{a}o Carlos, Universidade de S\~{a}o Paulo, \\
             Caixa Postal 369, 13560-970, S\~{a}o Carlos, SP, Brazil \\
             E-mail address: marcelo$\_$march@bol.com.br}
\date{\today}
%
\begin{abstract}
\vspace*{0.1mm}
\begin{center}
\rule[0.1in]{142mm}{0.4mm}
\end{center}
The Cahill-Glauber approach for quantum mechanics on phase-space is extended to the finite dimensional case through the use of discrete
coherent states. All properties and features of the continuous formalism are appropriately generalized. The continuum results are
promptly recovered as a limiting case. The Jacobi Theta functions are shown to have a prominent role in the context. \\
\vspace*{0.1mm}
\begin{center}
\rule[0.1in]{142mm}{0.4mm}
\end{center}
\end{abstract}
\maketitle
\section{Introduction}

The search for discrete quantum phase-space quasiprobability distribution functions is a subject of continuous and growing interest in
the literature \cite{wooters,gapi1,cohendet,gapi2,opat,voros,gama,zhang,luis,haki,muk,wooters2,vourdas}. The possibility of representing
quantum systems characterized by a finite-dimensional state space by such discrete quasidistributions lays the ground for interesting
developments and fruitful applications on quantum computation and quantum information theory \cite{r1s3,r2s3,r3s3,r4s3,r5s3,r6s3,r7s3,
r8s3}. It is well known that, as a well established counterpart to the discrete case, a huge variety of quasiprobabilty distribution
functions can be defined upon continuous phase-space \cite{lee}. In this sense, the Cahill-Glauber (CG) approach \cite{cahill} to the
subject has proved to be a powerful mapping technique that provides a general class of quasiprobability distribution functions, where
the Wigner, Glauber-Sudarshan and Husimi functions appear as particular cases. Therefore, it might be considered as a wide-range
phase-space approach to quantum mechanics regarding degrees of freedom with classical counterparts.

The aim of this paper is to present a discrete extension of the CG approach. Such extension is not obtained from that approach but,
instead, properly constructed out of the finite dimensional context. Furthermore, this {\sl ab initio} construction inherently embodies
the discrete analogues of the desired properties of the CG formalism. In particular, discrete Wigner, Husimi and Glauber-Sudarshan
quasiprobability distribution functions are obtained. Thus, besides the theoretical interest of its own, such extension has direct
applications in quantum information processing, quantum tomography and quantum teleportation, which are explored in a following work
\cite{nois}.

This work is organized as follows: In the next section we briefly outline the CG approach, setting the stage for section III, where our
proposal for a discrete extension of the CG mapping kernel is presented. In section IV basic properties of the mapping technique are
discussed, and the continuum limit is carried out on section V. Finally, section VI contains our summary and conclusions. Also,
important calculations are detailed in the Appendix.

\section{The Cahill-Glauber Mapping Kernel}

For the sake of clarity, in what follows we will briefly review the central ideas which constitute the core of the CG approach, and
that will be properly generalized in the following sections. Basically, the cornerstone of the formalism is the mapping kernel
(hereafter $\hbar = 1$)
\be
\lb{CG1}
{\bf T}^{(s)}(q,p) = \int_{-\infty}^{\infty} \frac{dp^{\prime} dq^{\prime}}{2\pi} \exp \lbk -i p^{\prime}(q-\mathbf{Q}) \rbk
\exp \lbk i q^{\prime} (p-\mathbf{P}) \rbk \exp \lpar - \frac{i}{2} p^{\prime} q^{\prime} \rpar \exp \lbk \frac{s}{4}(q^{\prime 2}
+ p^{\prime 2}) \rbk \; ,
\ee
which is responsible for the mapping of bounded operators on the continuous phase-space, being $s$ a complex variable satisfying the
condition $|s| \leq 1$. Here, the momentum and coordinate operators obey the Weyl-Heisenberg commutation relation $[ {\bf Q}, {\bf P}]
= i {\bf 1}$. Since the above expression explicitly depends on $s$, this parameter labels an infinite family of mapping kernels. Each
mapping kernel can be seen as the double Fourier transform of the displacement generators multiplied by a phase factor $\exp \lbk (i/2)
p^{\prime} q^{\prime} \rbk$ and by the folding function $\exp \lbk (s/4)(q^{\prime 2} + p^{\prime 2}) \rbk$. For purposes which will
become evident later, we write the mapping kernel as
\be
\lb{II}
{\bf T}^{(s)}(q,p) = \int_{-\infty}^{\infty} \frac{dp^{\prime} dq^{\prime}}{2 \pi} \exp \lbk -i p^{\prime}(q-\mathbf{Q}) \rbk
\exp \lbk i q^{\prime}(p-\mathbf{P}) \rbk \exp \lpar - \frac{i}{2} p^{\prime} q^{\prime} \rpar (\lg 0 | q^{\prime},p^{\prime}
\rg)^{-s} \; ,
\ee
where $| q^{\prime}, p^{\prime} \rg$ is a coherent state.

The mapping of a given operator is achieved by the trace operation $\mathcal{O}^{(s)}(q,p) = \tr [ {\bf T}^{(s)}(q,p) {\bf O}]$, being
$\mathcal{O}^{(s)}(q,p)$ the function which represents ${\bf O}$ in the associated usual phase-space. The mapping is one-to-one, and the
operator is reobtained from its associated function by
\bd
{\bf O} = \int_{-\infty}^{\infty} \frac{dpdq}{2\pi} \, \mathcal{O}^{(s)}(q,p) {\bf T}^{(-s)}(q,p) \; .
\ed
It is clear that, for each operator, there is an infinite family of associated functions labeled by $s$. In particular, the phase-space
representatives of the density operator are referred to as quasiprobability distributions functions and have, obviously,
distinguishable importance \cite{lee}. One of the great virtues of the Cahill-Glauber approach is that three special and important
types of quasidistributions, namely the Glauber-Sudarshan $(s=1)$, Wigner $(s=0)$ and Husimi functions $(s=-1)$, are particular cases.
Each of these functions have been extensively explored and reviewed in the literature \cite{schleich}.

A particular mapping kernel, characterized by a given parameter $s$, can be expressed in terms of another mapping kernel with a
different parameter value. The same holds true to the functions associated with a given operator. In fact, the procedure in the latter
case can be easily shown to be the same as in the former. That is, we may discuss only the relation between the mapping kernels,
knowing that equivalent relations are observed by the associated functions. In this way, the connection between the two mapping kernels
is seen to be given by the trace of the product
\br
\lb{fold0.5}
\tr [ {\bf T}^{(s_1)}(q_{1},p_{1}) {\bf T}^{(s_{2})}(q_{2},p_{2}) ] &=& \int_{-\infty}^{\infty} \frac{dq dp}{2\pi} \exp \lbr i \lbk
q (p_{1}-p_{2}) - p (q_{1}-q_{2}) \rbk \rbr \stackrel{(\lg 0 |q,p \rg)^{-(s_{1}+s_{2})}}{\overbrace{\exp \lbk \frac{s_{1}+s_{2}}{4}
( q^{2}+p^{2} ) \rbk}} \\
\lb{fold1}
&=& \frac{-2}{s_{1}+s_{2}} \exp \lbr \frac{2}{s_{1}+s_{2}} \lbk (p_{1}-p_{2})^{2} + (q_1-q_2)^{2} \rbk \rbr \qquad
\re (s_{1} + s_{2}) < 0 \; .
\er
We immediately recognize the important role played by the last exponential function in (\ref{fold0.5}), since, if the condition
$\re (s_{1}+s_{2}) < 0$ is not observed, the trace gives a divergent result. Thus, that condition imposes a constraint that defines a
hierarchy. That is, on continuous phase space there is a hierarchical structure of mapping kernels allowing one to express a given
phase-space function in terms of a Gaussian smoothing of another, and, as such, inverse relations do {\em not} exist. In other words,
{\em the Gaussian folding hierarchical structure observed by the quasidistributions has its roots in the functional form of} $\lg 0 |
q,p \rg$. We stress this particular point as the discrete equivalent to equation (\ref{fold0.5}) {\em does not} imply a hierarchical
relation.

\section{The discrete mapping kernel}
\subsection{Preliminaries}
\subsubsection{Operator bases}

Long ago Schwinger proposed the following set of operators to act as a basis on an operator space
\bd
{\bf S}(\eta,\xi) = \frac{1}{\sqrt{N}} {\bf U}^{\eta} {\bf V}^{\xi} \exp \lpar \frac{\pi i}{N} \eta \xi \rpar \; ,
\ed
where the ${\bf U}$'s and ${\bf V}$'s are the so-called Schwinger unitary operators \cite{schw}, $N$ is the dimension of the associated
state space and the indices $\{ \eta,\xi \}$ run on any complete set of residues mod$(N)$; in particular we choose the closed interval
$\lbk -\ell,\ell \rbk$, with $\ell = (N-1)/2$. For simplicity, we shall restrict ourselves to the odd $N$ case. Even dimensionalities,
for the purposes of this paper, can also be dealt with simply by working on non-symmetrized intervals.

The set $\{ {\bf S}(\eta,\xi) \}_{\eta,\xi = -\ell,\ldots,\ell}$ spans a complete and orthonormal basis on the $N^{2}$ space of linear
operators acting on finite complex vectorial spaces, in the sense that, as the trace operation stands as the inner product on operator
spaces, any linear operator can be written as
\be
\lb{1}
{\bf O} = \sum_{\eta,\xi = -\ell}^{\ell} \tr \lbk {\bf S}^{\dagger}(\eta,\xi) {\bf O} \rbk {\bf S}(\eta,\xi) \; .
\ee
The fundamental result
\bd
\tr \lbk {\bf S}^{\dagger}(\mu,\nu) {\bf S}(\eta,\xi) \rbk = \delta_{\eta,\mu}^{[N]} \, \delta_{\xi,\nu}^{[N]}
\ed
ensures that this decomposition is unique. The superscript $[N]$ on the Kroenecker deltas denotes that they are different from zero
whenever their indices are mod$(N)$ congruent. The Schwinger basis elements also obey the property \cite{gama}
\be
\label{2}
{\bf S}^{\dagger}(\eta,\xi) = {\bf S}(-\eta,-\xi) \; .
\ee

\subsubsection{Discrete Coherent States}

The Schwinger operator bases elements also act as displacement operators on a particular reference state to form discrete coherent
states as \cite{gapi2,gama}
\be
\lb{cohe}
| \eta,\xi \rg = \sqrt{N} {\bf S}(\eta,-\xi) |0,0 \rg \; ,
\ee
where the reference state is written by means of the Jacobi $\vartheta_{3}$-function (whose explicit form is shown in Appendix A) as
\be
\lb{vacuo}
|0,0 \rg = \frac{1}{\mathcal{N}} \sum_{\gamma = - \ell}^{\ell} \vartheta_{3} \lpar 2a \gamma | 2ia \rpar | u_{\gamma} \rg \; ,
\ee
where $\{ | u_{\gamma} \rg \}_{\gamma = - \ell,\ldots,\ell}$ are the eigenstates of the unitary operator ${\bf U}$,
\bd
{\mathcal{N}}^{2} = \frac{1}{2 \sqrt{a}} \lbk \vartheta_{3}(0|i a) \vartheta_{3}(0|4ia) + \vartheta_{4}(0|ia) \vartheta_{2}(0|4ia) \rbk
\ed
is the normalization constant, and $a=(2N)^{-1}$. Due to the properties of the $\vartheta_{3}$-function, the reference state above is
preserved under the action of the Fourier operator \cite{gama,mehta}
\bd
\mbox{\boldmath $\mathfrak{F}$} |0,0 \rg = |0,0 \rg \; ,
\ed
where
\bd
\mbox{\boldmath $\mathfrak{F}$} = \sum_{\gamma = -\ell}^{\ell} |v_{\gamma} \rg \lg u_{\gamma} | \; ,
\ed
and $\{ | v_{\gamma} \rg \}_{\gamma = - \ell,\ldots,\ell}$ are the eigenstates of ${\bf V}$, with $\lg u_{\mu} | v_{\gamma} \rg = \exp
[ (2 \pi i /N) \mu \gamma ]$. Parity of the $\vartheta_{3}$-function also ensures that $\lg u_{\kappa } | 0,0 \rg = \lg u_{-\kappa} |
0,0 \rg$, from which it follows
\be
\label{par}
\lg 0,0 | \mu,\nu \rg = \lg 0,0 | -\mu,-\nu \rg \; .
\ee

There are, of course, a number different recipes of discrete coherent states, some of them also in connection with $\vartheta$-functions, for instance \cite{zhang,voros}.

\subsection{The extended mapping kernel}

Now let us define the extended mapping kernel as
\bd
{\bf S}^{(s)}(\eta,\xi) = {\bf S}(\eta,\xi) \lbk \mathcal{K}(\eta,\xi) \rbk^{-s}
\ed
where $s$ is a complex number satisfying $\left| s \right| \leq 1$, and $\mathcal{K}(\eta,\xi)= \lg 0,0| \eta,\xi \rg$ denotes the
overlap of coherent states explicitly calculated in Appendix A. The set $\{ {\bf S}(\eta,\xi) \}_{\eta,\xi = -\ell,\ldots,\ell}$ itself
spans a complete and orthogonal basis on operator space. Nevertheless, we can go back to decomposition (\ref{1}), use equation
(\ref{2}), and introduce convenient factors to get
\bd
{\bf O} = \sum_{\eta,\xi = -\ell}^{\ell} \tr \lbk {\bf S}(-\eta,-\xi) {\bf O} \rbk {\bf S}(\eta,\xi)
{\underbrace{[ \mathcal{K}(-\eta,-\xi)]^{s} [\mathcal{K}(\eta,\xi)]^{-s}}_{1}} \; ,
\ed
where equation (\ref{par}) has been used. Conveniently grouping the terms the new decomposition reads
\be
\lb{novdec}
{\bf O} = \sum_{\eta,\xi = -\ell}^{\ell} \tr \lbk {\bf S}^{(-s)}(-\eta,-\xi) {\bf O} \rbk {\bf S}^{(s)}(\eta,\xi) \; .
\ee
Now, introducing the double Fourier transform of ${\bf S}^{(s)}(\eta,\xi)$, i.e.
\bd
{\bf T}^{(s)}(\eta,\xi) = \frac{1}{\sqrt{N}} \sum_{\mu,\nu = -\ell}^{\ell} {\bf S}^{(s)}(\eta,\xi) \exp \lbk - \frac{2\pi i}{N}
(\eta \mu + \xi \nu ) \rbk \; ,
\ed
and its Fourier inverse, we can, after a few steps, write equation (\ref{novdec}) as
\be
\lb{decomp}
{\bf O} = \frac{1}{N} \sum_{\mu,\nu = -\ell}^{\ell} \mathcal{O}^{(-s)}(\mu,\nu) {\bf T}^{(s)}(\mu,\nu) \; ,
\ee
with $\mathcal{O}^{(-s)}(\mu,\nu) = \tr \lbk {\bf T}^{(-s)}(\mu,\nu) {\bf O} \rbk$, defining a one-to-one mapping between operators and
functions defined on a discrete phase-space $\{\mu ,\nu \}$, where explicitly
\be
\lb{novabase}
{\bf T}^{(s)}(\mu,\nu) = \frac{1}{N} \sum_{\eta,\xi = -\ell}^{\ell} {\bf U}^{\eta} {\bf V}^{\xi} \exp \lbk -\frac{2\pi i}{N}
( \eta \mu + \xi \nu ) \rbk \exp \lpar \frac{\pi i}{N} \eta \xi \rpar [ \mathcal{K}(\eta,\xi) ]^{-s} \; ,
\ee
and $\mathcal{K}(\eta,\xi)$ can be shown to be a sum of products of Jacobi $\vartheta$-functions (as seen in Appendix A),
\br
\lb{fold}
\mathcal{K}(\eta,\xi) &=& \frac{1}{4 \sqrt{a} \mathcal{N}^{2}} \lbr \vartheta_{3} (a \eta | ia) \vartheta_{3} (a \xi | ia) +
\vartheta_{3} (a \eta | ia) \vartheta_{4} (a \xi | ia) \exp (i \pi \eta) \right. \nonumber \\
& & + \left. \vartheta_{4} (a \eta | ia) \vartheta_{3} (a \xi | ia) \exp (i \pi \xi) + \vartheta_{4} (a \eta | ia) \vartheta_{4}
(a \xi | ia) \exp \lbk i \pi (\eta + \xi + N) \rbk \rbr \; .
\er
The new kernel, written as in equation (\ref{novabase}), allows us to conclude that the above sum of products of $\vartheta$-functions
plays, in the discrete phase-space, the role reserved to the Gaussians in the continuous case.

\section{Properties}
\subsection{Basic general properties}

From the properties of the mapping kernel it is straightforward to obtain general properties of the associated functions in
phase-space. We observe that {\em all} following properties correctly generalize the continuous CG ones. First we note that
\be
\lb{prop1}
\textrm{(i)} \lbk {\bf T}^{(s)}(\mu,\nu) \rbk^{\dagger} = {\bf T}^{(s^{\ast})}(\mu,\nu) \; ,
\ee
implying that the mapping kernel is Hermitian for real values of the parameter $s$. As a direct consequence, the phase-space
representatives of Hermitian operators are real.

Direct calculations also show that
\br
\lb{prop2}
\textrm{(ii)} &\! \! \!& \frac{1}{N} \sum_{\mu,\nu = -\ell}^{\ell} {\bf T}^{(s)}(\mu,\nu) = {\bf 1} \; , \\
\textrm{(iii)} &\! \! \!& \tr \lbk {\bf T}^{(s)}(\mu,\nu) \rbk = 1 \; , \\
\textrm{(iv)} &\! \! \!& \tr \lbk {\bf T}^{(s)}(\mu,\nu) {\bf T}^{(-s)}(\mu^{\prime},\nu^{\prime}) \rbk = N
\delta_{\mu,\mu^{\prime}}^{[N]} \delta_{\nu,\nu^{\prime}}^{[N]} \; .
\er
The property (iv) is a crucial one from which expression (\ref{decomp}) could be immediately obtained. From this property also follows
the general result
\bd
\tr ( {\bf AB} ) = \frac{1}{N} \sum_{\mu,\nu = - \ell}^{\ell} \mathcal{A}^{(s)}(\mu,\nu) \mathcal{B}^{(-s)}(\mu,\nu) \; .
\ed
In fact, property (iv) is a particular case of the general expression
\be
\lb{gentr}
\tr \lbk {\bf T}^{(s)}(\mu,\nu) {\bf T}^{(t)}(\mu^{\prime},\nu^{\prime}) \rbk = \frac{1}{N} \sum_{\eta,\xi = - \ell}^{\ell}
\exp \lbr \frac{2 \pi i}{N} \lbk \eta (\mu^{\prime} - \mu) + \xi (\nu^{\prime} - \nu) \rbk \rbr [ \mathcal{K}(\eta,\xi) ]^{-(t+s)} \; ,
\ee
which is the counterpart of equation (\ref{fold0.5}). It must be stressed that this expression is {\em always} well defined, even for
$\re (t+s) < 0$, as $\mathcal{K}(\eta,\xi)\neq 0$.

\subsection{Particular cases}

There are three important particular cases to be discussed:
\begin{itemize}
\item[i)\hspace{.8 em}(s=0)] In such a case it is easy to see that
\be
\lb{DWW}
{\bf T}^{(0)}(\mu,\nu) = {\bf G}(\mu,\nu) \; ,
\ee
where ${\bf G}(\mu,\nu)$ is the mapping kernel introduced by Galetti and Piza, which is a discrete generalization of the Weyl-Wigner
mapping kernel \cite{gapi1,gapi2,ruga,ruga2}. In that case, being $\ro$ the density operator, equation (\ref{decomp}) would read, for
${\bf O} = \ro$,
\bd
\ro = \frac{1}{N} \sum_{\mu,\nu = -\ell}^{\ell} \mathcal{W}(\mu,\nu) {\bf G}(\mu,\nu) \; ,
\ed
with $\mathcal{W}(\mu,\nu) = \tr \lbk {\bf G}(\mu,\nu) \ro \rbk$ a discrete Wigner function.

\item[ii)\hspace{.8 em}(s=-1)] A fundamental property of our mapping kernel is
\be
\lb{fund}
{\bf T}^{(-1)}(\mu,\nu) = | \mu,\nu \rg \lg \mu,\nu | \; ,
\ee
which can be proved decomposing the coherent state projector in the Schwinger operator basis as (using equations (\ref{1}) and
(\ref{2}))
\bd
| \mu,\nu \rg \lg \mu,\nu | = \frac{1}{N} \sum_{\eta,\xi = -\ell}^{\ell} {\bf U}^{\eta} {\bf V}^{\xi} \exp \lpar \frac{i \pi}{N}
\eta \xi \rpar \tr \lbk | \mu,\nu \rg \lg \mu,\nu | {\bf V}^{-\xi} {\bf U}^{-\eta} \exp \lpar - \frac{i \pi}{N} \eta \xi \rpar \rbk
\ed
which, by applying the definition of the coherent states, equation (\ref{cohe}), reads
\bd
| \mu,\nu \rg \lg \mu,\nu | = \frac{1}{N} \sum_{\eta,\xi = -\ell}^{\ell} {\bf U}^{\eta} {\bf V}^{\xi} \lg 0,0 | {\bf V}^{\nu}
{\bf U}^{-\mu} {\bf V}^{-\xi} {\bf U}^{-\eta} {\bf U}^{\mu} {\bf V}^{-\nu} | 0,0 \rg \; ,
\ed
and using the Weyl commutation relation, ${\bf U}^{\alpha} {\bf V}^{\beta} = \exp [- (2 \pi i/N) \alpha \beta] {\bf V}^{\beta}
{\bf U}^{\alpha}$,
\bd
|\mu ,\nu \rangle \langle \mu ,\nu |=\frac{1}{N}\sum_{\eta ,\xi=-\ell}^{\ell}
\mathbf{U}^{\eta} \mathbf{V}^{\xi} \exp \left[ -\frac{2\pi i}{N}(\eta \mu +\xi \nu )
\right] \exp \lpar \frac{\pi i}{N}\eta \xi \rpar \langle 0,0|\eta ,\xi \rangle \; ,
\ed
where in the last step parity of $\lg 0,0 | \eta,\xi \rg$ with respect to $\eta$ was used. This proves our assertion. As a consequence,
the phase-space decomposition of the density operator, associated with this particular value of the parameter $s$, reads
\bd
\lb{glauber}
\ro = \frac{1}{N} \sum_{\mu,\nu = -\ell}^{\ell} P(\mu,\nu) | \mu,\nu \rg \lg \mu,\nu |
\ed
allowing us to identify $P(\mu ,\nu )$ as a discrete Glauber-Sudarshan distribution.

\item[iii)\hspace{.8 em} (s=1)] In this case we may write
\bd
\ro = \frac{1}{N} \sum_{\mu,\nu = -\ell}^{\ell} \tr \lbk {\bf T}^{(-1)}(\mu,\nu) \ro \rbk {\bf T}^{(1)}(\mu,\nu) \; ,
\ed
which is simply
\bd
\ro = \frac{1}{N} \sum_{\mu,\nu = -\ell}^{\ell} \mathcal{H}(\mu,\nu) {\bf T}^{(1)}(\mu,\nu) \; .
\ed
By definition $\mathcal{H}(\mu,\nu) = \lg \mu,\nu | \ro | \mu,\nu \rg$ is positive definite, and it can be identified as a discrete
Husimi function.
\end{itemize}

As any operator can be decomposed by the use of expression (\ref{decomp}), it follows that we are allowed to write
\bd
{\bf T}^{(-1)}(\mu,\nu) = \frac{1}{N} \sum_{\sigma,\lambda = -\ell}^{\ell} \tr \lbk {\bf T}^{(-0)}(\sigma,\lambda ) {\bf T}^{(-1)}
(\mu,\nu) \rbk {\bf T}^{(0)}(\sigma,\lambda) \; ,
\ed
where the minus signal was kept only for clarity. We then use equation (\ref{gentr}) to write explicitly
\be
\lb{ffold}
\tr \lbk {\bf T}^{(0)}(\sigma,\lambda) {\bf T}^{(-1)}(\mu,\nu) \rbk = \frac{1}{N} \sum_{\eta,\xi = -\ell}^{\ell} \exp \lbr
\frac{2 \pi i}{N} [\eta (\mu - \sigma) + \xi (\nu -\lambda)] \rbk \mathcal{K}(\eta,\xi) \; ,
\ee
that is, the discrete Fourier transform of the $\mathcal{K}(\eta,\xi)$ is the folding function. The above result can also be written in
the compact form
\bd
\tr \lbk {\bf T}^{(0)}(\sigma,\lambda) {\bf T}^{(-1)}(\mu,\nu) \rbk = \lg \mu,\nu | {\bf G}(\sigma,\lambda) | \mu,\nu \rg \; ,
\ed
which is precisely the Wigner function associated with a coherent state $| \mu,\nu \rg$. We therefore have
\be
\lb{hier1}
{\bf T}^{(-1)}(\mu,\nu) = \frac{1}{N} \sum_{\sigma,\lambda = -\ell}^{\ell} \lg \mu,\nu | {\bf G}(\sigma,\lambda) | \mu,\nu \rg
{\bf T}^{(0)}(\sigma,\lambda) \; .
\ee
In the same form we now decompose ${\bf T}^{(0)}(\mu,\nu)$ as
\bd
{\bf T}^{(0)}(\mu,\nu) = \frac{1}{N} \sum_{\sigma,\lambda = -\ell}^{\ell} \tr \lbk {\bf T}^{(0)}(\sigma,\lambda)
{\bf T}^{(-1)}(\mu,\nu) \rbk {\bf T}^{(1)}(\sigma,\lambda) \; ,
\ed
which allows us to use once again the above result for the trace and write
\be
\lb{hier2}
{\bf T}^{(0)}(\mu,\nu) = \frac{1}{N} \sum_{\sigma,\lambda = -\ell}^{\ell} \lg \mu,\nu | {\bf G}(\sigma,\lambda) | \mu,\nu \rg
{\bf T}^{(1)}(\sigma,\lambda) \; .
\ee
Multiplying both equations (\ref{hier1}) and (\ref{hier2}) by the density operator $\ro$ and taking the trace, we are led to the
suggestive results
\br
\mathcal{H}(\mu,\nu) &=& \frac{1}{N} \sum_{\sigma,\lambda = -\ell}^{\ell} \lg \mu,\nu | {\bf G}(\sigma,\lambda) | \mu,\nu \rg
\mathcal{W}(\sigma,\lambda) \; , \nn \\
\mathcal{W}(\mu,\nu) &=& \frac{1}{N} \sum_{\sigma,\lambda = -\ell}^{\ell} \lg \mu,\nu | {\bf G}(\sigma,\lambda) | \mu,\nu \rg
P(\sigma,\lambda) \; , \nn
\er
which are the discrete counterparts of the well known Gaussian smoothing that occurs in the continuous case, in agreement with the
hierarchy present in that context.

It must be stressed, however, that, opposed to the continuous case, it is now possible to write
\br
{\bf T}^{(0)}(\mu,\nu) &=& \frac{1}{N} \sum_{\sigma,\lambda = -\ell}^{\ell} \Lambda (\mu - \sigma, \nu - \lambda) {\bf T}^{(-1)}
(\sigma,\lambda) \nn \\
{\bf T}^{(1)}(\mu,\nu) &=& \frac{1}{N} \sum_{\sigma,\lambda = -\ell}^{\ell} \Lambda (\mu - \sigma, \nu -\lambda) {\bf T}^{(0)}
(\sigma,\lambda) \; , \nn
\er
where
\be
\lb{invfold}
\Lambda (\mu-\sigma,\nu-\lambda) = \frac{1}{N} \sum_{\eta,\xi = -\ell}^{\ell} \exp \lbr \frac{2 \pi i}{N} [\eta (\mu-\sigma)+
\xi (\nu-\lambda)] \rbr [ \mathcal{K}(\eta,\xi) ]^{-1} \; ,
\ee
which, at least in principle, can always be calculated (we remind again that $\mathcal{K}(\eta,\xi)$ is finite and different from zero).

Also very illustrative is the result that follows from the decomposition 
\be
\lb{fold2}
{\bf T}^{(-1)}(\mu,\nu) = \frac{1}{N} \sum_{\sigma,\lambda = -\ell}^{\ell} \tr \lbk {\bf T}^{(-1)}(\sigma,\lambda) {\bf T}^{(-1)}
(\mu,\nu) \rbk {\bf T}^{(1)}(\sigma,\lambda) \; .
\ee
With
\bd
\tr \lbk {\bf T}^{(-1)}(\sigma,\lambda) {\bf T}^{(-1)}(\mu,\nu) \rbk = \frac{1}{N} \sum_{\eta,\xi = -\ell}^{\ell} \exp \lbr
\frac{2 \pi i}{N} [\eta (\mu -\sigma) + \xi (\nu -\lambda) ] \rbr [\mathcal{K}(\eta,\xi)]^{2} \; ,
\ed
which can be shown to be $| \lg \mu,\nu | \sigma,\lambda \rg |^{2}$, we have
\be
\lb{fold3}
{\bf T}^{(-1)}(\mu,\nu) = \frac{1}{N} \sum_{\sigma,\lambda = -\ell}^{\ell} | \lg \mu,\nu | \sigma,\lambda \rg |^{2} \, {\bf T}^{(1)}
(\sigma,\lambda) \; .
\ee
Thus $| \lg \mu,\nu | \sigma,\lambda \rg |^{2}$ itself, which is the Husimi function associated with the discrete coherent-state
$| \mu,\nu \rg$, acts here as the smoothing function.

\section{Continuum limit}

Following the procedure detailed in both \cite{ruga,ruga2}, the continuum limit of the mapping kernel (\ref{novabase}) is reached as
follows: we introduce the scaling parameter $\eps = (2 \pi /N )^{1/2}$, which will become infinitesimal as $N \rightarrow \infty$, and
the two Hermitian operators
\be
\lb{29}
{\bf P} = \sum_{\mu = -\ell}^{\ell} \mu \epsilon p_{0} | v_{\mu} \rg \lg v_{\mu} | \qquad {\bf Q} = \sum_{\mu^{\prime} = -\ell}^{\ell}
\mu^{\prime} \epsilon q_{0} | u_{\mu^{\prime}} \rg \lg u_{\mu^{\prime}} | \; ,
\ee
constructed out of the projectors of the eigenstates of ${\bf U}$ and ${\bf V}$. The parameters $p_{0}$ and $q_{0}$, with $p_{0} q_{0}
= \hbar = 1$, are chosen to be real, carrying units of momentum and position, respectively, while $\epsilon p_{0}$ and $\epsilon q_{0}$
are the distance between successive eigenvalues of the ${\bf P}$ and ${\bf Q}$ operators. Then, rewriting the Schwinger operators as
\be
\lb{28}
{\bf V} = \exp \lpar \frac{i \epsilon {\bf P}}{p_{0}} \rpar \qquad {\bf U} = \exp \lpar \frac{i \epsilon {\bf Q}}{q_{0}} \rpar \; ,
\ee
and performing the change of variables $q^{\prime} = - q_{0} \epsilon \xi$, $p^{\prime} = p_{0} \epsilon \eta$, $p = p_{0} \eps \nu$
and $q = q_{0} \eps \mu$, we obtain
\bd
{\bf T}^{(s)}(q,p) = \sum_{q^{\prime} = - q_{0} \epsilon \ell}^{q_{0} \epsilon \ell} \sum_{p^{\prime} = - p_{0} \epsilon \ell }^{p_{0}
\epsilon \ell} \frac{\Delta q^{\prime} \Delta p^{\prime}}{2 \pi} \exp \lbk - i p^{\prime} (q-{\bf Q}) \rbk \exp \lbk i q^{\prime}
(p-{\bf P}) \rbk [\mathcal{K}(p^{\prime}/p_{0} \epsilon, - q^{\prime}/q_{0} \epsilon)]^{-s} \exp \lpar - \frac{i}{2} q^{\prime}
p^{\prime} \rpar \; .
\ed
As $N \rightarrow \infty$, it follows that $\Delta q^{\prime} \rightarrow dq^{\prime}$ and $\Delta p^{\prime} \rightarrow dp^{\prime}$.
Since the continuum limit of the discrete coherent-states has been already discussed in \cite{gapi2,gama}, it is clear that the term
$[ \mathcal{K}(p^{\prime}/p_{0} \epsilon, - q^{\prime}/q_{0} \epsilon)]^{-s}$, which is even, will go to $(\lg 0 |q^{\prime},p^{\prime}
\rg )^{-s}$. Therefore we end up with
\bd
{\bf T}^{(s)}(q,p) = \int_{-\infty}^{\infty} \frac{dp^{\prime} dq^{\prime}}{2 \pi} \exp \lbk - i p^{\prime}(q-{\bf Q}) \rbk \exp
\lbk i q^{\prime}(p-{\bf P}) \rbk \exp \lpar - \frac{i}{2} p^{\prime} q^{\prime} \rpar (\lg 0 | q^{\prime},p^{\prime} \rg )^{-s} \; ,
\ed
which is exactly the mapping kernel (\ref{II}) of Cahill and Glauber.

\section{Concluding Remarks}

The results obtained here show a genuine discrete mathematical structure which closely parallels the one of Cahill and Glauber. This
was achieved pursuing the lines proposed in \cite{gapi1}, which makes use of the discrete Fourier transform of the Schwinger operator
basis, to deal with the discrete phase-space problem. Now, we stress that expression (\ref{II}) is as simple as it is important, since
it clarifies the role of the coherent states overlap within the CG approach. By its turn, the discrete coherent states proposed in
\cite{gapi2,zhang,gama,voros} provide a natural path for a discrete extension of the CG formalism, while the properties of these states
have played a crucial role as they led, for example to the basic equation (\ref{fund}). Thus, the coherent states overlap can be seen
as the link between the discrete and continuous approaches. Furthermore, the continuum limit presented in section V ensures that the CG
mapping scheme is correctly recovered through a limiting procedure which is mathematically consistent \cite{ruga2,bar1,bar2,bar3}.

It is worth mentioning that Opatrn\'{y} {\em et al} \cite{opat} have pursued a goal similar to ours. Although both approaches share
virtues, our formalism presents mathematical features that allow us to achieve farther reaching results. It is precisely the correct
choice for the reference state, and the mathematical procedure adopted here, that lead to the obtention of such a wide set of important
results.

The use of Schwinger operators is crucial if one is concerned with the problem of ordering. As they are unitary shift operators,
equation (\ref{novabase}) makes it clear that the associated expansion is necessarily linked to a particular ordering of ${\bf U}$ and
${\bf V}$, which can be directly connected to the ${\bf Q}$ and ${\bf P}$ ordering of the continuous case.

Concerning the role of the Jacobi Theta functions in the discrete phase-space context -- they are implicit in the $\mathcal{K}
(\eta,\xi)$ term --, comparison with the usual CG results makes it evident that the Gaussian (or anti-Gaussian) terms, which are present
in the continuous case, are here replaced by the sum of products of $\vartheta$-functions (\ref{fold}), and its Fourier transform
(\ref{ffold}); both play here the role of the smoothing functions.

It is always important to emphasize the discrete case's peculiar features that do not have correspondence in the continuum. A plain
example of these is expressed by the well-behaved function given by equation (\ref{invfold}), whose continuum limit clearly diverges,
as the hierarchical structure presented in (\ref{fold1}) would imply. The finite character of the discrete scenario prevents such a
behaviour since, even if some terms in equation (\ref{invfold}) might become large for large $N$, they remain always finite due to the
behaviour of the $\vartheta$-functions. This allows one -- to give a extreme example -- to express, in the discrete scenario, the
Glauber-Sudarshan function in terms of the Husimi function.

Finally, it is worth mentioning that the mathematical formalism developed here opens new possibilities of investigations in quantum
tomography and quantum teleportation. These considerations are under
current research and will be published elsewhere (for instance, see reference \cite{nois}).

\section*{Acknowledgments}

This work has been supported by Funda\c{c}\~{a}o de Amparo \`{a} Pesquisa do Estado de S\~{a}o Paulo (FAPESP), Brazil, project nos.
03/13488-0 (MR), 01/11209-0 (MAM), and 00/15084-5 (MAM and MR). DG acknowledges partial financial support from the Conselho Nacional
de Desenvolvimento Cient\'{\i}fico e Tecnol\'{o}gico (CNPq), Brazil.

\appendix
\section{The discrete coherent-state overlap}

The vacuum associated with the discrete coherent-states (\ref{cohe}), given by equation (\ref{vacuo}), is written in terms of the Jacobi
$\vartheta_{3}$-function, which reads
\bd
\vartheta_{3} (a \mu | ia) = \sum_{\alpha = - \infty}^{\infty} \exp \lpar -a \pi \alpha^{2} + 2 \pi i a \mu \alpha \rpar
\ed
with $a=(2N)^{-1}$. Explicitly, then, the overlap $\lg \eta,\xi | \mu,\nu \rg$ is:
\br
\lg \eta,\xi | \mu,\nu \rg &=& \frac{1}{\mathcal{N}^{2}} \exp \lbk 2 \pi i a (-\mu \nu + \eta \xi) \rbk \lg 0,0 | {\bf V}^{\xi}
{\bf U}^{-\eta} {\bf U}^{\mu} {\bf V}^{-\nu} |0,0 \rg \nn \\
&=& \frac{1}{\mathcal{N}^{2}} \exp \lbr 2 \pi i a [- \mu \nu + \eta \xi + 2 \xi (\mu-\eta)] \rbr \sum_{\kappa,\nu = -\ell}^{\ell}
\vartheta_{3} (2a \kappa | 2ia) \vartheta_{3} (2a \nu | 2ia) \lg v_{\nu} | {\bf U}^{\mu-\eta} {\bf V}^{\xi -\nu} | v_{\kappa} \rg
\nn \; ,
\er
which, by evaluating the matrix element within the summation, gives
\bd
\frac{1}{\mathcal{N}^{2}} \exp \lbr 2 \pi i a [-\mu \nu + \eta \xi + 2 \xi (\mu-\eta )] \rbr \sum_{\kappa = -\ell}^{\ell}
\vartheta_{3} (2a \kappa | 2i a) \vartheta_{3} (2a(\kappa + \mu - \eta) | 2ia) \exp \lbk - \pi i a \kappa (\nu - \xi) \rbk \; .
\ed
Now let us call
\be
\lb{A}
A(\mu,\nu) = \sum_{\kappa = -\ell}^{\ell} \vartheta_{3} (2a \kappa | 2ia) \vartheta_{3} (2a (\kappa+\mu) | 2ia)
\exp (- \pi i a \kappa \nu) \; ,
\ee
so we write
\bd
\lg \eta,\xi | \mu,\nu \rg = \frac{1}{\mathcal{N}^{2}} \exp \lbr 2 \pi i a [- \mu \nu + \eta \xi + 2 \xi (\mu-\eta)] \rbr
A(\mu-\eta,\nu-\xi) \; .
\ed
The task now is to evaluate explicitly the term $A(\mu,\nu)$. Prior to that, some properties of $A(\mu,\nu)$ might be verified from the
start. First, we note that $A(\mu,\nu) = A(\nu,\mu)$ (using the fact that the $\vartheta_{3}$ is an eigenfunction of the discrete
Fourier transform), and also $A(\mu + N \kappa,\nu + N \kappa^{\prime}) = A(\mu,\nu)$, for $\kappa$ and $\kappa^{\prime}$ integers.

Let us now evaluate $A(\mu,\nu)$. We start using the $\vartheta_{3}$ definition to get
\bd
A(\mu,\nu) = \sum_{\kappa = -\ell}^{\ell} \sum_{\alpha,\beta = -\infty}^{\infty} \exp (-4 \pi i a \kappa \mu) \exp (-2 \pi a \alpha^{2}
+ 4 \pi i a \kappa \alpha) \exp \lbk -2 \pi a \beta^{2} + 4 \pi i a (\kappa + \nu) \beta \rbk \; .
\ed
The sum over $\kappa$ can be readily carried out, which gives us $N \delta_{\beta,\mu-\alpha}^{[N]}$. Thus, $\beta$ will assume the
values $\mu - \alpha + N \gamma$, where $\gamma$ is an arbitrary integer, yielding
\bd
N \sum_{\alpha,\gamma = -\infty}^{\infty} \exp (-2 \pi a \alpha^{2}) \exp \lbk -2 \pi a (\mu - \alpha + N \gamma)^{2} + 4 \pi i a \nu
(\mu - \alpha + N \gamma) \rbk \; .
\ed
The last term in the second exponential is equal to one, and we shift the sum over $\alpha$ by $\mu$ in order to get
\bd
N \sum_{\alpha,\gamma = -\infty}^{\infty} \exp \lbk -4 \pi a \lpar \alpha - \frac{\gamma N}{2} \rpar^{2} - \frac{\pi \gamma^{2} N}{2} -
2 \pi a \mu^{2} - 4 \pi i a \alpha (\nu-i\mu) \rbk \; .
\ed
Now, we split the sum over $\gamma$ in contributions coming from the even (e) and odd (o) integers as follows: $A = A_{e} + A_{o}$.
Consequently, the even term can be dealt with by shifting the sum over $\alpha$ by $-N \gamma$,
\br
A_{e}(\mu,\nu) &=& N \sum_{\alpha,\gamma = - \infty}^{\infty} \exp \lbk -4 \pi a \alpha^{2} - 2 \pi \gamma^{2} N - 2 \pi a \mu^{2} + 4
\pi i a (\alpha + N \gamma)(\nu -i \mu) \rbk \nn \\
&=& N \sum_{\alpha = -\infty}^{\infty} \exp \lbk -4 \pi a \alpha ^{2} + 4 \pi i a \alpha (\nu -i\mu) \rbk
\sum_{\gamma = - \infty}^{\infty} \exp \lbk -2 \pi \gamma^{2} N + 4 \pi i a \gamma N (\nu -i \mu) \rbk \nn \\
&=& N \vartheta_{3} (2a (\nu - i \mu) | 4ia) \, \vartheta_{3} (i \mu | 2iN) \exp \lpar -2 \pi a \mu^{2} \rpar \; , \nn
\er
and, in a similar fashion, the odd term gives
\bd
A_{o}(\mu,\nu) = N \vartheta_{3} (2a [\nu -i (\mu+N)] | 4ia) \, \vartheta_{2} (i \mu | 2iN) \exp \lpar - \frac{\pi N}{2} - \pi \mu -
2 \pi a \mu^{2} \rpar \; .
\ed

Next, using a relation (all the following relations for $\vartheta$-functions come from \cite{bellman})
\bd
\vartheta_{3} \lpar \varsigma + \left. \frac{\tau}{2} \right| \tau \rpar = \exp \lpar -i \pi \frac{\tau}{4} - i \pi \varsigma \rpar
\vartheta_{2} (\varsigma | \tau)
\ed
we get
\bd
A_{o}(\mu,\nu) = N \vartheta_{3} (2a [\nu -i (\mu+N)] | 4ia) \, \vartheta_{3} (i(\mu+N) | 2iN) \exp \lbk -2 \pi a (\mu+N)^2 \rbk \; .
\ed
Then
\br
A(\mu,\nu) &=& N \vartheta_{3} (2a[\nu -i\mu ] | 4ia) \, \vartheta_{3} (i \mu | 2iN) \exp \lpar -2a \pi \mu^{2} \rpar \nn \\
& & + \, N \vartheta_{3} (2a[\nu -i(\mu +N)] | 4ia) \, \vartheta_{3} (i(\mu+N) |2iN) \exp \lbk -2 \pi a (\mu +N)^{2} \rbk \; . \nn
\er
Now we recall the fundamental property of $\vartheta$-functions,
\bd
\vartheta_{3} \lpar \left. \frac{\varsigma}{i\tau} \right| \frac{i}{\tau} \rpar = \sqrt{\tau} \exp \lpar \frac{\pi \varsigma^{2}}{\tau}
\rpar \vartheta_{3} (\varsigma | i \tau) \; ,
\ed
which applied to the $\vartheta_{3} (i \mu | 2iN)$ and $\vartheta_{3} (i (\mu +N) | 2iN)$ terms above gives
\br
A(\mu,\nu) &=& \sqrt{\frac{N}{2}} \, \vartheta_{3} (2a (\nu - i \mu) | 4ia) \, \vartheta_{3} (a \mu | ia) \exp \lpar - \pi a \mu^{2}
\rpar \nn \\
& & + \, \sqrt{\frac{N}{2}} \, \vartheta_{3} (2a [\nu - i(\mu +N)] | 4ia) \, \vartheta_{3} (a (\mu +N) | ia) \exp \lbk - \pi a
(\mu + N)^{2} \rbk \; . \nn
\er
Still, one has to use the relation
\bd
\vartheta_{3} (\varsigma | \tau) = \frac{1}{2} \lbk \vartheta_{3} \lpar \left. \frac{\varsigma}{2} \right| \frac{\tau}{4} \rpar +
\vartheta_{4} \lpar \left. \frac{\varsigma}{2} \right| \frac{\tau}{4} \rpar \rbk
\ed
in the first $\vartheta$-function of each of the two terms above to finally employ the relations of quasi-periodicity
\br
\vartheta_{3} (\varsigma + m \tau | \tau) &=& \exp \lpar - \pi i \tau m^{2} + 2 \pi im \varsigma \rpar \vartheta_{3} (\varsigma | \tau)
\nn \\
\vartheta_{4} (\varsigma + m \tau | \tau) &=& \exp \lpar - \pi i \tau m^{2} + 2 \pi im \varsigma + \pi im \rpar \vartheta_{4}
(\varsigma | \tau) \; , \nn
\er
to reach the desired result
\br
A(\mu,\nu) &=& \frac{1}{4 \sqrt{a}} \exp (2 \pi ia \mu \nu) \lbr \vartheta_{3} (a \mu | ia) \vartheta_{3} (a \nu | ia) + \vartheta_{3}
(a \mu | ia) \vartheta_{4} (a \nu | ia) \exp (i \pi \mu) \right. \nn \\
&& + \left. \vartheta_{4} (a \mu | ia) \vartheta_{3} (a \nu | ia) \exp (i \pi \nu) + \vartheta_{4} (a \mu | ia) \vartheta_{4}
(a \nu | ia) \exp [ i \pi (\nu +\mu + N) ] \rbr \; . \nn
\er
This completes the calculation of $\lg \eta,\xi | \mu,\nu \rg$.

Finally, it is convenient to write down the result for the particular case $\mathcal{K}(\mu,\nu) \equiv \lg 0,0 | \mu,\nu \rg$, i.e.,
\be
\lb{eqdoA}
\mathcal{K}(\mu,\nu) = \frac{1}{\mathcal{N}^{2}} \exp \lpar - 2 \pi i a \mu \nu \rpar A(\mu,\nu) \; .
\ee
%

\end{document}